\documentclass[a4paper,twoside,10pt]{article}

\input{jgrg20.sty}

\usepackage{amsfonts}
\usepackage{amsmath,amssymb}
\usepackage[T1]{fontenc}
\usepackage{mathrsfs}
\usepackage{epsfig,epstopdf}

\usepackage[absolute]{textpos}

\usepackage{graphics,wrapfig,times}
\usepackage{graphicx}

\usepackage{color}
\definecolor{D}{rgb}{0.00,0.17,0.48}
\definecolor{M}{rgb}{0.00,0.02,0.83}
\definecolor{L}{rgb}{0.58,0.79,1.00}
\definecolor{R}{rgb}{1,0,0}
\definecolor{B}{rgb}{0.00,0.00,0.00}
\definecolor{P}{rgb}{0.00,0.30,0.60}
\definecolor{W}{rgb}{1,1,1}

\TPGrid[40mm,40mm]{15}{25}  

 \makeatletter
 \newcommand{\pback}[1]{{
   \let\@rrow=\leftarrowfill
   \mathchoice{\AIN@stemPullBack{#1}{\@rrow}}{\AIN@stemPullBack{#1}{\@rrow}}
     {\AIN@indxPullBack{#1}{\@rrow}}{\AIN@indxPullBack{#1}{\@rrow}}}
   \vphantom{#1}}

 \newcommand{\AIN@stemPullBack}[2]{
   \vtop{\mathsurround=0pt
   \ialign{##\crcr$\textstyle{#1}\strut$\crcr
     \noalign{\kern-0.4ex\nointerlineskip}{\tiny#2}\crcr}}}

 \newcommand{\AIN@indxPullBack}[2]{
   \vtop{\mathsurround=0pt
   \ialign{##\crcr\hfil$\scriptstyle{#1}$\hfil\crcr
     \noalign{\kern+0.4ex\nointerlineskip}{\tiny#2}\crcr}}}

\def\bar{\overline}
\def\be{\begin{equation}}
\def\ee{\end{equation}}
\def\bea{\begin{eqnarray}}
\def\eea{\end{eqnarray}}
\def\ba{\begin{array}}
\def\ea{\end{array}}

\def\={\hateq}
\def\puto#1{\rlap{\raise.5ex\hbox{\char'27}}{#1}}

\newcommand{\nn}{\nonumber}
\newcommand{\half}{\frac{1}{2}}
\def\eth{\text{\dh}}
\def\thorn{\text{\th}}
\def\a{\alpha}
\def\b{\beta}
\def\c{\gamma}

\def\.{\cdot}

\def\Re{{\rm Re}}
\def\Im{{\rm Im}}
\def\l{\ell}

\def\be{\begin{equation}}
\def\ee{\end{equation}}
\def\bea{\begin{eqnarray}}
\def\eea{\end{eqnarray}}
\def\ba{\begin{array}}
\def\ea{\end{array}}

\newcommand{\eqhat}{\mathrel{\widehat\mathalpha{=}}}
\def\={\eqhat}

\begin{document}
%
\pagestyle{fancy}
\fancyhead{}
  \fancyhead[RO,LE]{\thepage}
  \fancyhead[LO]{Y. H. Wu}          
  \fancyhead[RE]{JGRG20 Proceedings}    
\rfoot{}
\cfoot{}
\lfoot{}
\label{P81}                             
\title{%
  Gravitational radiation and angular momentum flux from a spinning dynamical black hole
}
%
\author{%
  Yu-Huei Wu\footnote{Email address: yhwu@mail.phy.ncu.edu.tw}$^{(a)}$$^{(b)}$
  and
  Chih-Hung Wang\footnote{Email address: chwang1101@phys.sinica.edu.tw, chwang@phy.ncu.edu.tw}$^{(b)}$$^{(c)}$$^{(d)}$
  }
%
\address{%
  $^{(a)}$Center for Mathematics and Theoretical Physics, National Central University, Chungli, 320, Taiwan.\\
  $^{(b)}$Department of Physics, National Central University, Chungli, 320, Taiwan.\\
  $^{(c)}$Department of Physics, Tamkang University, Tamsui, Taipei 25137, Taiwan.\\
  $^{(d)}$Institute of Physics, Academia Sinica, Taipei 115, Taiwan.}
%
\abstract{
A four-dimensional asymptotic expansion scheme is used to study the next order effects of the nonlinearity near a spinning dynamical black hole. The angular momentum flux and energy flux formula are then obtained by asymptotic expansion and the compatibility of the coupling Newman-Penrose equations. After constructing the reference frame in terms of the compatible constant spinors, the energy-momentum flux is derived and it is related to the black hole area growth. Directly from the flux formula of the spinning dynamical horizon, we find that the physically reasonable condition on the positivity of the gravitational energy flux yields that the shear will monotonically decrease with time.
}

\section{Introduction}

In this paper, we use the Bondi-type coordinates to write the null tetrad for a spinning dynamical horizon (DH). The boundary conditions for the quasi-local horizons can be expressed in terms of Newman-Penrose (NP) coefficients from the Ashtekar's definition on DH.
Unlike Ashtekar \textit{et al}'s \cite{Ashtekar99b, Ashtekar02} three dimensional analysis, we adopts a 4-dimensional asymptotic expansion to study the neighborhoods of generic isolated horizons (IHs) and dynamical horizons (DHs). Since the asymptotic expansion has been used to study gravitational radiations near the null infinity \cite{NU, ENP}, it offers a useful scheme to analyze gravitational radiations approaching another boundary of space-time, black hole horizons.
We first set up a null frame with the proper gauge choices near quasi-local horizons and then expand Newman-Penrose (NP) coefficients, Weyl, and Ricci curvature with respect to radius. Their fall-off can be determined from NP equations, Bianchi equations, and exact solutions, e.g., the Vaidya solution.  From the reduction and the decoupling of the equations governing the Weyl scalars, instead of assuming $\Psi_0,\Psi_1=0$ on DH, we set $\Psi_1, \Psi_3$ vanishing on a spinning DH. This serves as a \textit{peeling property} for a spinning DH and is a similar setting with the perturbation method (Also see Chandrasekhar \cite{Chandrasekhar}). This approach allows one to see the next order contributions from the nonlinearity of the full theory for the quasi-local horizons.
%
We have shown that the quasi-local energy-momentum flux formula for a non-rotating DH by using asymptotic expansion yields the same result as Ashtekar-Krishman flux \cite{Wu2007, WuWang-PRD-2009}. For slow rotating DH, we have presented our results in \cite{Wu2007}, however, it makes too many assumptions to be satisfactory. Furthermore, the flux formula has a shear (NP coefficient $\sigma$) and a angular momentum (NP coefficient $\pi$) coupling term. Since it is unclear whether the existence of this term carries any physical meaning or it may due to our assumptions, we thereby study the fast spinning case and extend our previous work on IHs and DHs into a more general case.
It is known that DH is a three-dimensional spacelike hypersurface with its topology $R^1 \times S^2$, so the shape of its 2-dimentioanl cross section will depend on the gauge freedom of choosing foliation. One may image that there exists a gauge choice which can make DH's cross section looks like a 2-sphere. 
The existence of angular momentum will not change the boundary condition for the null infinity, however, it will affect the boundary conditions of  a black hole.
In GR, quasi-local mass expressions for Kerr solution disagree one another \cite{Bergqvist92}. Different quasi-local expressions give different values of quasi-local mass for Kerr black hole. At null infinity, there is no generally accepted definition for angular momentum \cite{Szabados-04}. 
%
By the aid of using asymptotic constant spinor to define spin frame as the reference frame for our observation, mass and angular momentum flux can be calculated. 

\section{Angular momentum and energy-momentum and their flux of a spinning DH} \label{J-DH}


We choose the incoming null tetrad $n_a=\nabla_a v$ to be the gradient of
the null hypersurface $v=const.$ We then have $g^{ab} v_{,a} v_{,a}
=0$. It gives us the gauge conditions $\nu=\mu-\bar\mu=
\gamma+\bar\gamma=\bar\a+\b -\bar\pi=0$. Then we further choose $n^a$ flag plane parallel, it implies $\c=0$. For the setting of outgoing null tetrad $\l$, we first choose $\l$ to be a geodesic and use null rotation type III to make $\epsilon-\bar\epsilon=0$.  We choose $m,
\bar m$ tangent to the cross section $S$, and thus $\rho\hat{=}\bar\rho, \pi\hat{=}\bar
\tau$. In the comoving coordinate $(v, r', x^2, x^3 )$ we have
 $\l^a=(1, U - \dot R_\Delta, X^2, X^3)$,$ n^a=(0,-1,0,0),$
          $m^a=(0,0, \xi^2,\xi^3)$, 
where $\dot R_\Delta(v)$ is the rate of changing effective radius of DH and $r'=r-R_\Delta(v)$ and $R_\Delta(v)$ is effective radius of a spinning DH.
%
%
Since we use $\kappa=\nu=0, \sigma\neq 0, \lambda\neq 0$, therefore one can set $\Psi_1\hat{=} \Psi_3\hat{=}0$ as \textit{peeling properties for a spinning DH}. This is a  similar with perturbation method and one may refer to p. 175 and p. 180 in \cite{Chandrasekhar}. The falloff of the Weyl scalars is
algebraically general (this is a more general setting than \cite{Wu2007} and \cite{WuWang-PRD-2009}) on DH where
\be\ba{l} \Psi_1=\Psi_3=O(r'), \Psi_0=\Psi_2=\Psi_4 =O(1).
\ea\ee
By considering Vaidya solution as our compared basis for matter field part, the falloff of the Ricci spinor components are
\bea \Phi_{00}=O(1),\Phi_{22}=\Phi_{11}=\Phi_{02}=\Phi_{01}=\Phi_{21}=O(r').\eea

We adopt a similar idea of Bramson's asymptotic
frame alignment for null infinity \cite{Bramson75a} and apply it
to set up spinor frames for a spinning DH. We find the compatible conditions for spin frame
are
\bea \thorn_0 \lambda^0_0 &=& 0,
\eth_0 \lambda^0_0 + \sigma_0\lambda_1^0=0,
\eth_0 \lambda^0_1 - \mu_0 \lambda^0_0= 0,
\thorn_0 \lambda^0_1= - \bar\eth_0\lambda^0_0. \label{spinorCD}
 \eea
%
%
Since there are gauge freedoms on choosing the foliation of $DH$, we then assume that there exists a natural foliation to make DH cross section $S$ as a two sphere. Therefore, on a sphere with effective horizon radius $R_\Delta(v)$, one can set
$ \mu_0= -\frac{1}{R_\Delta}.$
Let $P, \mu_0$ on a sphere with radius $R_\Delta$, then $P \propto \frac{1}{R_\Delta}$. 
Moreover, the effective surface gravity is $\tilde{\kappa}=2\epsilon_0 = \frac{1}{2R_\Delta}$, and then $\mu_0=-4\epsilon_0$.
Check the commutation relation $[\delta_0,D_0]\lambda_0$ and $[\delta_0,D_0]\sigma_0$, it implies
$\ddot R_\Delta=0.$
This means that the horizon radius will not accelerate (no inflation). The dynamical horizon will increase with a constant speed.
After applying these conditions, we list the main equations that will be used later (for detail, see \cite{Wu-Wang-10})
\be\ba{lll}
%
 \textbf{(NR1)} &\dot R_\Delta [-\half(\Psi^0_2 +\bar\Psi^0_2)+\half(\eth_0\pi_0+ \bar\eth_0\bar\pi_0) -
\pi_0\bar\pi_0 ] =  \Phi_{00}^0,\\
\textbf{(NR2)} & \dot \sigma_0= \dot R_\Delta [-\eth_0
\bar\pi_0 -\bar\pi_0^2+ \sigma_0 \mu_0]+ 2 \epsilon_0 \sigma_0 +\Psi_0^0, \\
\textbf{(NR4)} & \frac{\bar\sigma_0 \ddot{R}_0 -\dot R_\Delta\dot{\bar\sigma_0}}{(\dot R_\Delta)^2} =\dot R_\Delta \Psi_4^0  +\bar\eth_0\pi_0+\pi_0^2
 + 2 \frac{\bar\sigma_0 \epsilon_0}{\dot R_\Delta}- \mu_0\bar\sigma_0,\\
\textbf{(NR7)} & \Re\Psi_2^0 = 2 \mu_0 \epsilon_0 -\pi_0\bar\pi_0-\Re\eth_0\pi_0,  \Im \Psi_2^0 =-\Im\eth_0\pi_0,\\
\textbf{(NB2)} & \dot \Psi_2^0=\dot R_\Delta[3\mu_0-\sigma_0\Psi_4^0]+\frac{\bar\sigma_0\Psi_0^0}{\dot R_\Delta}+\mu_0\Phi_{00}^0,\\
\textbf{(NR1)+(NR7)} & 2\Re\eth_0\pi_0=\frac{\Phi_{00}^0}{\dot R_\Delta}+2\mu_0\epsilon_0 ,\\
 %
%
  \textbf{(NR8)} &  \bar\eth_0\sigma_0 =0, 
 \textbf{(NR9)} \; -2\bar\sigma_0\bar\pi_0= \dot R_\Delta \mu_0\pi_0 ,
 \sigma_0\bar\eth_0\pi_0=-\half\dot R_\Delta \mu_0\bar\eth_0\bar\pi_0,\\
 \textbf{(NR3)}&  \dot\pi_0 = 2\dot R_\Delta \mu_0\pi_0, \textbf{(NR5)} \; \dot \a_0 =\dot R_\Delta\a_0\mu_0-\bar\sigma_0\bar\pi_0,
 \textbf{(NR6)} \; \dot \b_0 =\dot R_\Delta \mu_0(\bar\pi_0+\b_0) +\sigma_0 \pi_0.\nn \ea\ee

We use an asymptotically rotating Killing vector $\phi^a$ for a spinning DH.  It coincides with a rotating vector $\phi^\a\hat{=}\psi^a$ on a DH and is divergent free. It implies $\Delta_a \phi^a := S_{a0}^a S_a^{b0}\nabla_{b0} \phi^{a0}=0$. Therefore,
$ \bar m_a \delta \phi^a = -m_a\bar\delta\phi^a. $
Let $\phi^a=A m^a+ B \bar m^a$, we get $A=-B$. Therefore, it exists a function $f$ such that
$ \phi^a = \bar\delta f m^a-\delta f \bar m^a,$
which is type $(0,0)$. Since $f$ is type $(0,0)$, therefore $\delta f=\eth f$. By using Komar integral, the quasi-local angular momentum on a spinning DH is
\bea  J(R_\Delta) &=& -\frac{1}{4\pi} \oint_S  f \Im \Psi_2^0 d S_\Delta.\eea
From (NB2), we get $\Im \dot\Psi_2^0 = 3 \frac{\dot{R}_\Delta}{R_\Delta} \Im \eth_0 \pi_0 =-3 \frac{\dot{R}_\Delta}{R_\Delta} \Im \Psi_2^0$. Together with
$\frac{\partial}{\partial v} d S_\Delta= 2\frac{\dot{R}_\Delta}{R_\Delta} d S_\Delta$, the angular momentum flux for a spinning DH is
\bea \dot J(R_\Delta) = -\frac{1}{4\pi} \oint_S (\dot f -\frac{\dot R_\Delta}{R_\Delta} f) \Im \Psi_2^0 d S_\Delta
= -\frac{1}{4\pi} \oint_S \Im[ (\eth_0\dot f -\frac{\dot R_\Delta}{R_\Delta} \eth_0 f) \pi_0] d S_\Delta. \eea
We note that from $\frac{d}{d v}$ (NR7), it yields the same result. Here if $\pi_0\neq 0$ and $f(v,\theta,\phi)=G(\theta,\phi)R_\Delta(v)$, then $\dot J(R_\Delta)=0$. It then returns to the stationary case. If $\pi_0 =0$, i.e., $\Im\Psi_2^0=0$, then $J$ and $\dot J=0$. It then returns to the non-rotating black hole.

%
%
%


By using the compatible constant spinor conditions for a spinning dynamical
horizon (\ref{spinorCD}) and the results of the asymptotic
expansion, we get the quasi-local energy-momentum integral on a spinning
 dynamical horizon
%
\be\ba{llll} I(R_\Delta)&=&
-\frac{1}{4\pi}\oint \mu_0
\lambda_{0}^0
\bar\lambda_{0'}^0 d S_\Delta &\textrm{(i)}\\
&=&- \frac{1}{4\pi} \oint
\frac{\Re}{2\epsilon_0}[\Psi^0_2 +\delta_0 \pi_0+2\b_0\pi_0] \lambda_0^0 \bar\lambda^0_{0'} d S_\Delta. &\textrm{(ii)}\nn
\ea\ee

In order to calculate flux we need the time related condition (\ref{spinorCD}) of constant spinor of
dynamical horizon  and re-scale it.
Then $\dot{\lambda}^0_0=0$.
It's tedious but straightforward to calculate the flux expression.
It largely depends on the non-radial NP equations and the second
order NP coefficients. By using the results on two sphere foliation, we substitute them back into the energy-momentum flux formula to
simplify our expression.

\textbf{From (i):} Apply time derivative to (i), and then we obtain the \textit{quasi-local
energy momentum flux for dynamical horizon}
\bea \dot I(R_\Delta) = \frac{1}{4\pi}\oint \dot \mu_0
\lambda_{0}^0 \bar\lambda_{0'}^0 d S_\Delta. \label{NewsDH} \eea
where it is always \textit{positive}. Here $\dot\mu_0$ is the
\textit{news function of DH} that always has mass gain.
 Integrate the above equation with
respect to $v$ and use the results on two sphere,
we then have \cite{Wu-Wang-10}
\bea d I(R_\Delta) =\frac{1}{8\pi} \int \;^{(2)}R \lambda_{0'}^0
\bar\lambda_0^0 d S_\Delta d R_\Delta. \label{areaDH}\eea

\textbf{From (ii):}
The total energy momentum flux $F_{\textrm{total}}$  is equal to matter flux plus gravitational flux $F_{\textrm{total}}=F_{\textrm{matter}}+F_{\textrm{grav}}$.  We can write the gravitational flux equal to the shear flux plus angular momentum flux
$ F_{\textrm{grav}}=F_{\sigma}+F_{J}.$
 The coupling of the shear $\sigma_0$ and $\pi_0$ can be transform into $\pi_0$ terms by using (NR9), then we obtain
 \bea d I(R_\Delta) &=& \frac{1}{8\pi}\int \frac{ R_\Delta}{\dot R_\Delta}\{\frac{1}{\dot R_\Delta} \Phi_{00}^0 -2\frac{\sigma_0\bar\sigma_0}{\dot R_\Delta}( \frac{\partial}{\partial v} \ln( R_\Delta^2 \sigma_0\bar\sigma_0))  +3\frac{\dot R_\Delta\pi_0\bar\pi_0}{ R_\Delta}\} \lambda_{0}^0 \bar\lambda_{0'}^0 d
S_\Delta d R_\Delta. \label{ch6-dI}\eea
%
where $d v = \frac{d R_\Delta}{\dot R_\Delta}$. Here we note that if one wants to observe positive shear flux $-\frac{\partial}{\partial v}\ln (R_\Delta^2\sigma_0\bar\sigma_0)\geq 0$, it implies that
$ \dot \sigma_0 \leq 0,$
where $\dot R_\Delta, R_\Delta >0$ have been considered. So the shear on a spinning DH is monotonically decreasing with respect to $v$.

Recall that the total flux of Ashtekar-Krishnan \cite{Ashtekar02}, we
compare our expression with Ashtekar's expression.
%
%
If we choose $N= \lambda_{0}^0 \bar\lambda_{0'}^0$,
 then (\ref{ch6-dI}) together with (\ref{areaDH})
gives
%
\bea d I(R_\Delta) &=& \frac{1}{8\pi} \int \;^{(2)}R N d S_\Delta
d R_\Delta = \frac{1}{8\pi}\int \frac{ R_\Delta}{\dot R_\Delta}\{\frac{1}{\dot R_\Delta} \Phi_{00}^0 +2 k \frac{\sigma_0\bar\sigma_0}{\dot R_\Delta}
 +3\frac{\dot R_\Delta\pi_0\bar\pi_0}{ R_\Delta}\} N d
S_\Delta d R_\Delta  \label{DHflux} \eea 
where we define $\frac{\partial}{\partial v} \ln(R_\Delta^2\sigma_0\bar\sigma_0):=-k$ for the convenience. 
%
In the special case  $\frac{\partial}{\partial v} \ln(R_\Delta^2\sigma_0\bar\sigma_0):=-k$ where $k$ is a constant, we then have
\bea R_\Delta^2\sigma_0\bar\sigma_0 = A e^{-k v}.\eea
If $k>0$, $\sigma_0\searrow$. If $k<0$, $\sigma_0\nearrow$. Therefore, if we want to get positive gravitational flux, the shear $\sigma_0$ must decrease with time $v$ and $k>0$. On the contrary, the negative gravitational flux implies the shear must grow with time.  The negative mass loss from shear flux will make the dynamical horizon grow with time is physically  unreasonable.
Therefore, the shear flux  should be positive. This says that the shear on a spinning DH will decay to zero when time $v$ goes to infinity and the amount of shear flux $F_\sigma$ is finite.
\bea \sigma_0 \rightarrow 0, |v|\rightarrow \infty.  \eea
 Hence the dynamical horizon will settle down to an equilibrium state, i.e., isolated horizon.

\textbf{Laws of black hole dynamics}

LHS of eq. (\ref{DHflux}) can be written as
$\frac{d I(R_\Delta)}{2}= \frac{\tilde{\kappa}}{8\pi} d A = \frac{d R_\Delta}{2}$
where $A=4\pi R_\Delta^2$. For a time evolute vector $t^a=N \l^a -\Omega \phi^a$, the difference of horizon energy $d E^t$ can be calculated as follow
\bea d E^t &=& \frac{1}{16\pi} \int \frac{ R_\Delta}{\dot R_\Delta}\{\frac{1}{\dot R_\Delta} \Phi_{00}^0 +2 k \frac{\sigma_0\bar\sigma_0}{\dot R_\Delta}
 \} N +[3 N \pi_0\bar\pi_0-4\frac{\Omega}{\dot R_\Delta}\Im[(\eth_0\dot f-\frac{\dot R_\Delta}{R_\Delta}\eth_0 f )\pi_0] d V\nn\eea
and the generalized black hole first law for a spinning dynamical
horizon is
\bea \frac{\tilde{\kappa}}{8\pi} d A +\Omega d J = d E^t.\eea

\section{Conclusions}


We use a different peeling property from our earlier work \cite{WuWang-PRD-2009}\cite{Wu2007}. This leads to a physical picture that captures a non-spherical symmetric collapse of a spinning star and formation of dynamical horizon that finally settle down to an isolated horizon. Further from the peeling property, if the shear flux is positive, it excludes the possibility for a spinning DH to absorb the gravitational radiation from nearby gravitational sources. The mass and momentum are carried in by
the incoming gravitational wave and cross into dynamical horizon. We shall see that though it may exist outgoing wave on horizon, however, it will not change the boundary condition or make the contribution to the energy flux. A dynamical horizon forms inside the star and eat up all the incoming wave when it reaches the equilibrium state, i.e., isolated horizon.  The NP equations are simplified by transforming to a new comoving coordinate and a foliation of 2-sphere for a spinning DH. The existence about how to make such a choice of the foliation will be a further mathematical problem.
By using the compatibility of the coupling NP equations and the asymptotic constant spinors, the energy flux that cross into a spinning DH should be positive. The mass gain of a spinning DH can be quantitatively written as matter flux, shear flux and angular momentum flux. Further, a result  come out that the shear flux must be positive implies the shear must monotonically decay with respect to time. This is physically reasonable since black hole cannot eat infinite amount of gravitational energy when there is no other gravitational source near a spinning DH. We further found that the mass and mass flux based on Komar integral can yield the same result \cite{Wu-Wang-10}. Therefore, our results are unlikely expression dependent. For other quasi-local expressions remain the open question for the future study. It would be also interesting if one can generalize to a binary problem.


\end{document}